\documentclass[aps,pra,reprint,superscriptaddress,showpacs]{revtex4-1}

\usepackage{graphicx}
\usepackage{amsmath, amsthm, amssymb}
\usepackage{textcomp}

\usepackage{xcolor}

\begin{document}

\title{Anderson localisation in laser kicked molecules}

\author{Johannes Flo\ss}
\affiliation{Department of Chemical Physics, The Weizmann Institute of Science, 234 Herzl Street, Rehovot 76100, Israel}
\author{Shmuel Fishman}
\affiliation{Department of Physics, Technion - Israel Institute of Technology, Haifa 32000, Israel}
\author{Ilya Sh. Averbukh}
\affiliation{Department of Chemical Physics, The Weizmann Institute of Sciences, 234 Herzl Street, Rehovot 76100, Israel}

\date{\today}

\begin{abstract}
The paper  explores the prospects of observing the phenomenon of dynamical Anderson localisation  via non-resonant Raman-type rotational excitation of molecules by periodic trains of short laser pulses. We define conditions for such an experiment, and show that current femtosecond technology used for non-adiabatic laser alignment of linear molecules is sufficient for this task. Several observables which can serve as indicator for Anderson localisation are suggested for measurement, and the influence of experimental limitations imposed by laser intensity noise, finite pulse duration, limited number of pulses in a train, and thermal effects is analysed.
\end{abstract}



\maketitle



\section{Introduction}

Albeit of its simplicity, the periodically kicked rotor has attracted much attention in the recent decades.
One of the major reasons for the interest in this system is the research of quantum chaos.
In the classical regime, the periodically kicked rotor can exhibit chaotic motion, leading to an unbounded growth of the angular momentum.
A quantum mechanical rotor shows the chaotic behaviour for a limited period of time.
Eventually, the discreteness of the rotor energy leads to at least quasi-periodic motion and therefore a suppression of the diffusive growth of the angular momentum~\cite{casati79,fishman96}.
It was shown~\cite{fishman82,*grempel84} that this quantum suppression is due to a mechanism closely related to the Anderson localisation of electronic wave functions in disordered solids~\cite{anderson58}.
Destructive interferences lead to an exponential localisation of the wave function.
Another distinct feature of the quantum kicked rotor is the effect of quantum resonance~\cite{casati79,izrailev80}.
If a rotor is kicked at a period that is a rational multiple of the rotational revival time~\cite{revival,robinett04}, its energy grows quadratically with the number of kicks.

Linear molecules are a basic example of a quantum rotor.
It is consequential that an early proposal~\cite{bluemel86} for experiments on the quantum kicked rotor suggested using diatomic molecules kicked by a pulsed electric field.
It was proposed~\cite{bluemel86} to use a combination of several harmonics of a microwave field to create a train of microwave pulses, which would then interact with polar diatomic molecules like CsI.
This scheme (using a rotor with a permanent dipole moment) has been then analysed in many theoretical works during the last two decades~\cite{gong01a,shapiro07,lee06,spanner04}.
However, to the best of our knowledge, no experiment along these lines has been done yet (probably, due to the complexity of the required field source).

A different experimental approach to the kicked rotor problem was introduced by Raizen and co-workers~\cite{moore95,raizen99} who used a substitute system of ultracold atoms interacting with a pulsed standing light wave.
This system has become the standard set-up for observing effects of the $\delta$-kicked rotor, including quantum resonance and dynamical localisation~\cite{moore95} or the effects of noise on dynamical localisation~\cite{klappauf98,amman98}.
However, the non-discrete character of the atomic momentum complicates the observation of certain phenomena like quantum resonances and chaos assisted tunnelling.
To some degree, this was overcome by using a very narrow initial momentum distribution~\cite{steck01,hensinger01,ryu06}.
Periodically kicked molecules circumvent this problem, since the quantisation of the angular momentum ensures discreteness of the energy spectrum.

Recently~\cite{floss12}, we drew attention to the fact that current technology used for laser alignment of non-polar molecules offers an alternative for exploring the dynamics of the periodically kicked quantum rotor in a molecular system (see refs.~\cite{ourreview,ohshimareview} for a recent review of laser molecular alignment, and ref.~\cite{stapelfeldt03} for earlier studies). Here, the laser field affects the molecular rotation via Raman-type interaction~\cite{zon75,friedrich95,*friedrich95b}. The electric field of the pulse induces anisotropic molecular polarisation, interacts with it, and tends to align the molecular axis along the laser polarisation direction.
An ultra-short laser pulse acts like a kick, and the alignment is observed under field-free conditions after the pulse is over~\cite{ortigoso99,seideman99a,underwood05,3Dexp}.
In a recent experiment, a periodic train of eight pulses was used for inducing enhanced molecular alignment by repeated kicking under the condition of exact quantum resonance~\cite{cryan09}.
A direct experimental observation of the quantum resonance in periodically kicked molecules was achieved recently in~\cite{zhdanovich12a} by employing laser pulse trains with a variable period.
In this paper we extend our previous theoretical studies on periodically kicked molecules and elaborate in  detail on the prospects of observing Anderson localisation via non-resonant Raman-type rotational excitation by short laser pulses.

This paper is structured as follows.
In Sec.~\ref{sec.model} we introduce our model for the laser-molecule interaction and shortly review the connection between the dynamics of a periodically kicked molecule and the phenomenon of Anderson localisation.
In this section we also describe our numerical methods.
In Sec.~\ref{sec.ideal} we demonstrate different manifestations of the Anderson localisation phenomenon in the considered system.
These effects include  exponential localisation of the angular momentum distribution,  suppression of energy diffusion, and  finite survival probability of the initial state.
Sec.~\ref{sec.limitations} is devoted to experimental limitations and how they influence the prospected observations.
Finally, in Sec.~\ref{sec.discussion} we discuss the results and conclude.



\section{\label{sec.model}Model and calculation methods}

\subsection{Model}

\begin{figure}
\includegraphics{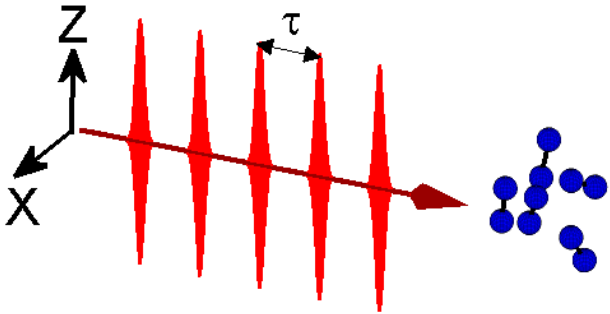}
\caption{\label{fig.model}(Colour online) The considered model: A train of ultrashort linearly polarised laser pulses interacts with linear molecules.
The time-delay $\tau$ between subsequent pulses is constant.}
\end{figure}

We consider interaction of linear molecules with a periodic train of linearly polarised laser pulses, as depicted in Fig.~\ref{fig.model}.
The laser pulses interact with a molecule via its electric polarisability.
The pulse's electric field induces a dipole and subsequently interacts with it.
After averaging over the fast oscillations of the electric field, the interaction potential for a single pulse is given as
\begin{equation}
V(\theta,t)=-\frac{1}{4}\left(\Delta\alpha\cos^2\theta+\alpha_{\perp}\right)\mathcal{E}^2(t) \,.
\label{eq.interaction}
\end{equation}
Here, $\theta$ is the angle between the molecular axis and laser polarisation direction, $\mathcal{E}(t)=\mathcal{E}_0 \exp[-t^2/(2\sigma^2)]$ is the envelope of the electric field, where $\sigma$ is a measure for the pulse duration, and $\Delta\alpha=\alpha_{\parallel}-\alpha_{\perp}$ is the polarisability anisotropy, where $\alpha_{\parallel}$ and $\alpha_{\perp}$ are the molecular polarisabilities along and perpendicular to the molecular axis, respectively.
We drop the second term in the brackets, as it has no angular dependence and therefore does not influence the rotational dynamics.
In the following, we express energy in units of $2B$ and time in units of $\hbar/(2B)$.
Here, $B=\hbar^2/(2I)$ is the rotational constant of the molecule, with $I$ being its moment of inertia.
Also we introduce the dimensionless interaction strength $P=\Delta\alpha/(4\hbar)\int\mathcal{E}^2(t)$, which corresponds to the typical angular momentum (in units of $\hbar$) transferred by the laser pulse to the molecule.
With these units, the Hamiltonian describing our system is given as
\begin{equation}
\hat H=\frac{\hat J^2}{2} - \frac{P}{\sqrt{\pi}\sigma}\cos^2\theta \sum_{n=0}^{N-1} \exp\left[(t-n\tau)^2/\sigma^2\right] \,,
\label{eq.hamiltonian}
\end{equation}
where $\hat J$ is the angular momentum operator, $\tau$ is the period of the pulse train and $N$ is the number of pulses.

\subsection{Mapping to a tight-binding model}

We will now shortly review how the mechanism of Anderson localisation is related to the localisation in angular momentum space of a periodically kicked linear molecule.
The connection between the kicked rotor and Anderson localisation was first established by Fishman, Grempel and Prange~\cite{fishman82,*grempel84}, and later on extended to kicked linear molecules~\cite{bluemel86}.
Here, we closely follow these references.

In order to show the relationship between the two phenomena, it is convenient to describe the dynamics in terms of the quasi-energy states~\cite{zeldovich67} $|\chi_{\alpha}\rangle$ (also called Floquet states) and the quasi-energies $\omega_{\alpha}$.
The quasi-energy states are solutions of the periodic time-dependent Sch\"odinger equation, which reproduce themselves (up to a phase factor) after every period of the field:
\begin{equation}
|\chi_{\alpha}(t+\tau)\rangle=e^{-i\omega_{\alpha}\tau}|\chi_{\alpha}(t)\rangle \,.
\label{eq.qestates}
\end{equation}
The quasi-energy states may  be expressed as
\begin{equation}
|\chi_{\alpha}(t)\rangle=e^{-i\omega_{\alpha}t} |u_{\alpha}(t)\rangle \,,
\end{equation}
where $|u_{\alpha}(t)\rangle=|u_{\alpha}(t+\tau)\rangle$ is a periodic function.
The quasi-energy states are assumed to form a complete basis, and the wave function in terms of the quasi-energy states is given as
\begin{equation}
|\Psi(t)\rangle=\sum_{\alpha}e^{-i\omega_{\alpha}t} | u_{\alpha}(t)\rangle \langle u_{\alpha} (0)|\Psi(0)\rangle \,.
\label{eq.expansionU}
\end{equation}
Here, $|\Psi(0)\rangle$ is the initial state.
Note that the expansion coefficients $\langle u_{\alpha} (0)|\Psi(0)\rangle$ are time-independent~\cite{zeldovich67}.

In order to show the connection to Anderson localisation, we have to look at the one-cycle propagator $\hat U(t+\tau,t)$, which propagates the wave function over one period.
The propagator can be expressed as
\begin{equation}
\hat U(t+\tau,t)=e^{-i\hat J^2 \tau/2} \mathcal{T} \exp\left[-i\int_t^{t+\tau} \mathrm{d}t' \tilde V(t',t)\right]\,,
\label{eq.evolutionoperator}
\end{equation}
where $\mathcal{T}$ is the time-ordering operator and
\begin{equation}
\tilde V(t',t)=e^{i\hat J^2(t'-t)/2}V(t')e^{-i\hat J^2(t'-t)/2} \,.
\end{equation}
With a suitable Hermitian operator $\hat W$, the propagator can be expressed as~\cite{bluemel86}
\begin{equation}
\hat U(t+\tau,t)=e^{-i\hat J^2 \tau/2} \frac{1+i\hat W}{1-i \hat W}\,.
\label{eq.UwithW}
\end{equation}
In the case of $\delta$-pulses, $W(\theta)$ is given as~\cite{bluemel86}
\begin{equation}
W(\theta)=-\tan\left[V(\theta)/2\right] \,.
\end{equation}

The quasi-energy states are related to the one-cycle propagator $\hat U(t+\tau,t)$ via
\begin{equation}
\hat U(t+\tau,t)|\chi_{\alpha}\rangle=e^{-i\omega_{\alpha}\tau}|\chi_{\alpha}\rangle \,.
\label{eq.psiinchi}
\end{equation}
By the use of Eq.~\eqref{eq.UwithW}, one can express the one-cycle evolution~\eqref{eq.psiinchi} as (compare with~\cite{fishman82,*grempel84,bluemel86})
\begin{equation}
T_{J}^{(\alpha)}u_{J}^{(\alpha,M)} + \sum_{J'}W_{J,J'}^{(M)}u_{J'}^{(\alpha,M)} =0 \,,
\label{eq.anderson}
\end{equation}
where
\begin{subequations}
\begin{align}
u_{J}^{(\alpha,M)}=&\langle J,M|\frac{1}{1-i\hat W}|u_{\alpha}\rangle \\
T_{J}^{(\alpha)}=&\tan\left(\tau\frac{\omega_{\alpha}-E_J}{2}\right) \label{eq.TJ}\\
W_{J,J'}^{(M)}=&\langle J,M|\hat W |J',M\rangle \,.
\end{align}
\end{subequations}
Here, $|J,M\rangle$ are the spherical harmonics, the eigenfunctions of a free rotor, and $E_J=J(J+1)/2$ are the corresponding eigenvalues (neglecting the centrifugal distortion term).
Note that the interaction~\eqref{eq.interaction} leaves the quantum number $M$ unchanged, so the latter becomes a mere parameter defined by the initial conditions.
Eq.~\eqref{eq.anderson} displays the problem of the periodically kicked molecule in the form of a tight-binding model, and therefore establishes the connection between localisation in the periodically kicked rotor and the Anderson model of localisation in disordered solids.

\subsection{Numerical Method}

For numerical purposes, it is convenient to work in the basis of the spherical harmonics, $|J,M\rangle$.
Since the interaction does not change $M$, we can treat it as a parameter.

The matrix elements of the one-cycle propagation operator $\mathbf{U}$ are obtained as follows.
First, we expand the wave function in the spherical harmonics,
\begin{equation}
|\Psi^{(M)}(t)\rangle=\sum_J C_J^{(M)} (t) e^{-iE_Jt} |J,M\rangle \,.
\label{eq.expansion}
\end{equation}
Inserting this expansion into the time-dependent Schr\"odinger equation with the Hamiltonian~\eqref{eq.hamiltonian}, we obtain a set of coupled differential equations for the expansion coefficients:
\begin{multline}
\frac{\partial C_{J}^{(M)}(t)}{\partial t}=
i \frac{P}{\sqrt{\pi}\sigma} \sum_{n=0}^{N-1} \exp\left[(t-n\tau-\tau/2)^2/\sigma^2\right]\\
\times\sum_{J'} C_{J'}^{(M)}(t) e^{-i(E_{J'}-E_{J})t} \langle J,M|\cos^2\theta |J',M\rangle \,.
\label{eq.differential}
\end{multline}
Here, we chose the timing of the pulses such that they are in the middle of the cycle.
The matrix elements of $\mathbf{U}$ are obtained by solving~\eqref{eq.differential} over one cycle (including a single pulse).
In particular, the element $U_{J,J'}^{(M)}$ is given as $C_{J}^{(M)}(t_0+\tau)e^{-iE_J\tau}$ with the initial conditions $C_{J}^{(M)}(t_0)=\delta_{J,J'}$, with $t_0=0$.

The quasi-energy states and the quasi-energies are obtained numerically as eigenstates and eigenvalues of $\mathbf{U}$ (see Eq.~\eqref{eq.psiinchi}).
The wave function after $N$ pulses can either be obtained by solving Eqs.~\eqref{eq.differential} for the whole pulse train, or by multiplying the initial state $N$ times by $\mathbf{U}$.
Which method is better suited depends mainly on the number of pulses and the pulse duration.

\subsection{\label{sec.dependenceonperiod}Dependence of the dynamics on the period}

\begin{figure}
\includegraphics{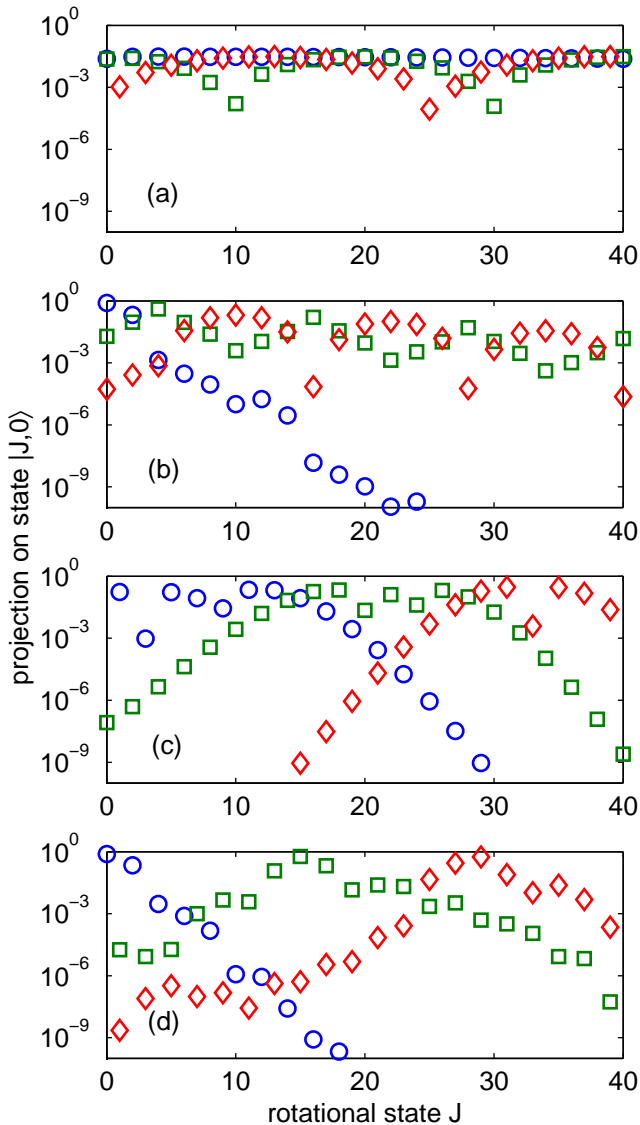}
\caption{\label{fig.qestates}(Colour online) Absolute square of the projection of selected quasienergy states on the angular momentum states $|J,0\rangle$.
Shown is the case of $\delta$-pulses with $P=3$.
(a) Full quantum resonance, $\tau=2\pi$. Under this condition, all quasienergy-states are extended over the full $J$-space.
(b) Fractional quantum resonance, $\tau=\pi/3$. Most states are extended over the full $J$-space, but some are localised close to $J=0$.
(c) Slight detuning form full quantum resonance, $\tau=2\pi+0.01$. The states are extended over several $J$-states.
(d) Far detuned from any resonance, $\tau=1$. All states are exponentially localised.}
\end{figure}

The dynamics of the kicked rotor depends strongly on the term $T_J^{(\alpha)}$ from Eq.~\eqref{eq.anderson}.
In particular, there are four regimes.
Firstly, $T_J^{(\alpha)}$ can be identical for every $J$.
This situation is called quantum resonance~\cite{casati79,izrailev80}, and is achieved if the period $\tau$ is an integer multiple of $2\pi$. (Note that the quantum resonance exists exactly only if the molecular centrifugal distortion is neglected.)
Under the condition of quantum resonance, all quasi-energy states are extended over the full angular momentum space, as is shown in Fig.~\ref{fig.qestates}~(a).
Secondly, $T_J^{(\alpha)}$ can be a periodic series in $J$.
This case is called fractional quantum resonance.
It occurs, if the kicking period is not an integer, but just a rational multiple of $2\pi$, $\tau=2\pi q/r$.
Here, most states are still extended over the full range of $J$-space, but there are few states which are exponentially localised close to $J=0$ (see Fig.~\ref{fig.qestates}~(b)).
These localised states may be regarded as ``surface states''.
They are due to a ``surface effect'' of the tight-binding model~\eqref{eq.anderson}, since $J=0$ is a lower limit for $J$.
The third case is if $T_J^{(\alpha)}$ is nearly periodic over a limited range of $J$.
This effect occurs if $\tau$ is slightly detuned from the resonant values, $\tau=2\pi+\epsilon$.
As a result, the quasi-energy states are extended over a limited range of $J$, as shown in Fig.~\ref{fig.qestates}~(c).
The last case, with which we will mainly deal in this paper, is if $T_J^{(\alpha)}$ is a random series in $J$.
In this case, the tight-binding model~\eqref{eq.anderson} exhibits  the phenomenon of Anderson localisation~\cite{anderson58}.
All states are exponentially localised in the angular momentum space, as shown in Fig.~\ref{fig.qestates}~(d).
In particular, at each site $J$ there is one state localised, although it may occur that two states mix and then occupy the same sites~\cite{bluemel86}.

In general, following the expansion~\eqref{eq.expansionU}, the wave function shows the same characteristics of extension or localisation in the momentum space as the quasi-energy states.
If the quasi-energy states are spread over the whole momentum space under the condition of quantum resonance, also the wave function will spread over the whole momentum space.
Vice versa, it will be localised like the quasi-energy states under the condition of Anderson localisation.
A special case are fractional resonances.
If the wave function is initially localised close to $J=0$, the localised ``surface states'' will have a significant overlap with the initial wave function, and therefore part of the wave function will remain localised to $J=0$, whilst another part will spread in momentum.

\begin{figure}
\includegraphics{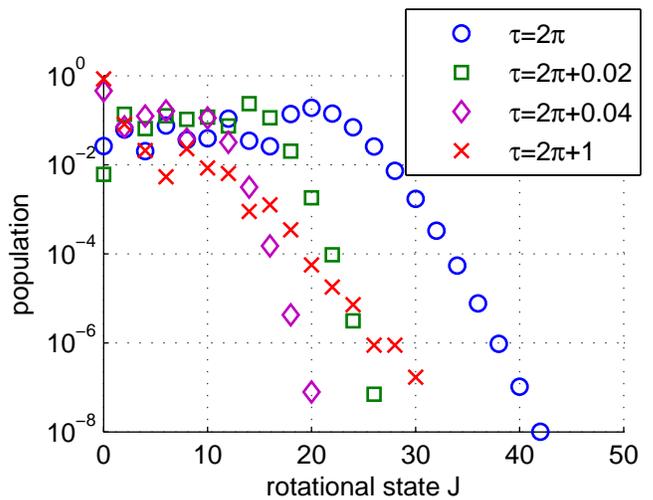}
\caption{\label{fig.population}(Colour online) Final population of the angular momentum states of a quantum rotor, after it was kicked by a train of eight pulses with $P=3$, for different pulse train periods.
The rotor was initially in the rotational ground state.}
\end{figure}

To finish the introductory section, we show the actual population distribution of the wave function for the above three different cases of the  train period, using a train of a finite number of pulses.
In Fig.~\ref{fig.population} the population of the rotational levels $J$ after a train of eight pulses is shown, for different train periods $\tau$.
Due to the selection rules of the Raman-type excitation ($J = 0, \pm 2$), the subsets of even and odd rotational levels evolve independently, so for clarity we show only the even ones.
The population of the odd levels behaves similarly.
Under the condition of exact resonance ($\tau=2\pi$, blue circles), the distribution is divided into a flat (in logarithmic scale) plateau region, and a fast decay after some cut-off value of $J$. The cut-off marks the maximum angular momentum supplied by $N$ pulses to a classical rotor (for a rigid rotor this is $J = NP$), and the fast decay is due to the "tunneling" into the classically forbidden region.
When the detuning is increased, one can see a monotonous deformation of the population curve (green squares and purple diamonds), while the general shape (plateau and fast decay after a cut-off) remains intact.
A more rigorous analysis of this close-to-resonance region can be found in~\cite{wimberger03,*wimberger04,floss12}.
For larger detuning (red crosses), the distribution is completely different and reflects Anderson localisation in the system.
Instead of the plateau, a clear linear (in logarithmic scale) decay over several orders of magnitude is seen, starting from $J = 0$.
In the remainder of this work, we will only consider this latter situation.



\section{\label{sec.ideal}Anderson localisation in a linear rigid molecule}

Anderson localisation is characterized by eigenfunctions that are exponentially localised in space, and consequently, the motion is confined in the vicinity of the initial position.
Here we consider Anderson localisation in the angular momentum space~\cite{fishman82,grempel84,bluemel86,floss12}.
As outlined in the following, Anderson localisation is reflected in several observables amendable to direct experimental observation.

In this Section, we consider three manifestations of Anderson localisation -- exponential localisation in angular momentum space, suppression of energy diffusion, and finite survival probability of the initial state -- and show that they can be observed for a periodically kicked molecule with current laser technology.
Due to the exponential localisation of the quasi-energy states -- discussed in Sec.~\ref{sec.dependenceonperiod} --, the population distribution of the angular momentum states is exponentially localised at the initial state.
The localisation length is a property of the time-evolution operator~\eqref{eq.evolutionoperator}, and can therefore be controlled by experimental parameters.
Another manifestation of the Anderson localisation is the suppression of energy diffusion.
Whilst a periodically kicked classical rotor undergoes a diffusive growth of the angular momentum (for sufficiently strong kicks), for a quantum rotor the localisation of the quasienergy states in angular momentum space suppresses this diffusive growth after few kicks.
When observing the absorbed rotational energy, one can therefore see an initial diffusive growth, which later changes to an oscillatory pattern.
As a third observable we consider the survival probability of the initial state.
If there is localisation, it remains finite.

In our simulations we include the effects of amplitude noise.
Anderson localisation is due to coherent effects and, in principle, is destroyed by an arbitrarily weak noise~\cite{guarneri84,ott84}.
In particular, for sufficiently strong noise, the phase coherence is destroyed such that the classical diffusion in the phase space is recovered~\cite{guarneri84,ott84,milner00}.
Therefore, noise can be used as a test to rule out localisation mechanisms which are not affected by noise, like adiabatic localisation (see Section~\ref{sec.limitations}).
On the other hand, noise can also be a problem in experiments by preventing the observation of localisation.
In our simulations, we demonstrate the effects of noise by introducing random Gaussian variations in the interaction strength.

In this section, we consider a simplistic model of a rigid linear molecule at zero temperature, kicked by ideal $\delta$-pulses.
In particular, this means that the initial state for the calculations is the rotational ground state $|0,0\rangle$, and the rotational levels are given as $E_J=J(J+1)/2$ (in units of $2B$).
The effects of deviations from this idealised case are considered in Section~\ref{sec.limitations}.
In the case of noisy pulse trains, the shown results are the average of 50 different pulse train realisations.

\begin{figure}
\includegraphics{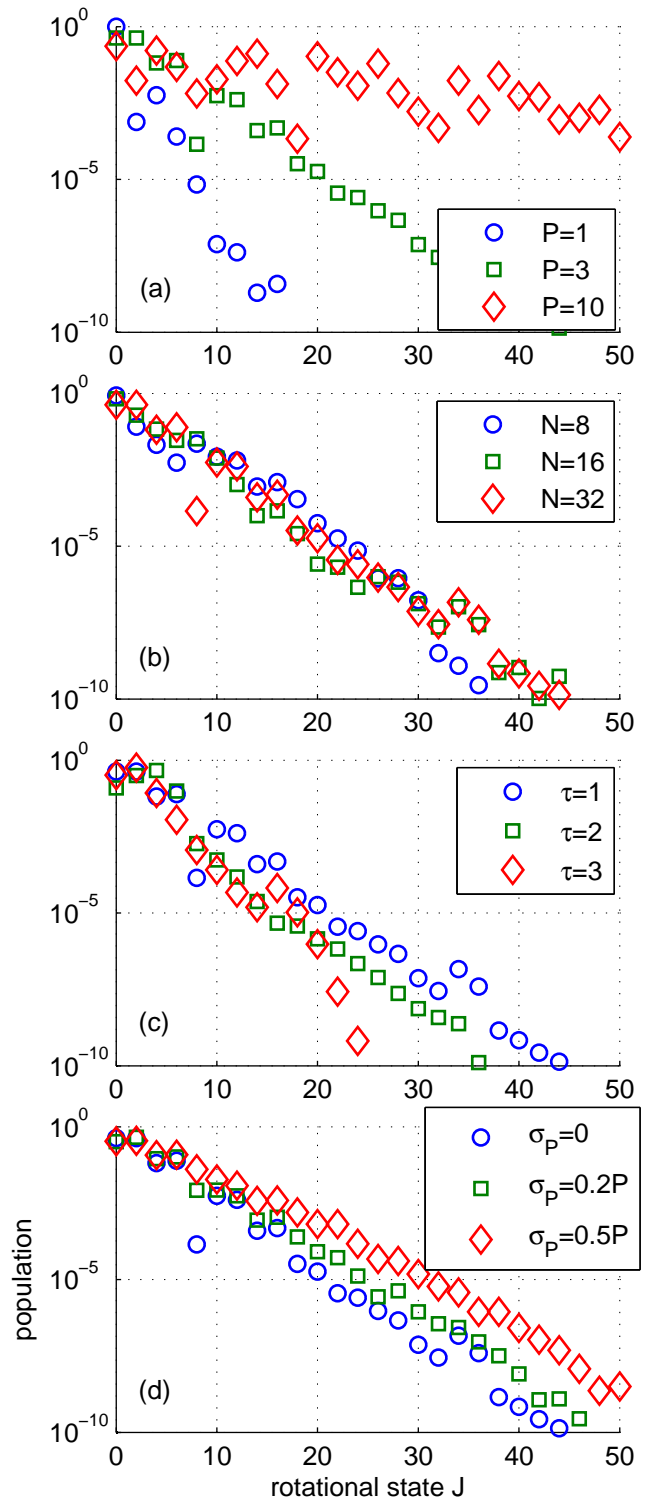}
\caption{\label{fig.exponential}(Colour online) Population of the even rotational states after a train of $\delta$-pulses.
In each panel, one parameter is varied: interaction strength $P$ (a), number of pulses $N$ (b), pulse train period $\tau$ (c), standard deviation $\sigma_P$ of the interaction strength (d).
The non-varied parameters are $P=3$, $N=32$, $\tau=1$, and $\sigma_P=0$.
}
\end{figure}

The mechanism of Anderson localisation causes the quasienergy states to be exponentially localised in momentum space.
As a result, the population distribution of the angular momentum states is exponentially localised around the initial state.
The localisation length is a property of the time-evolution operator $\hat U$ (see Eq.~\eqref{eq.evolutionoperator}), and is therefore independent from the number of pulses applied.
In the regime we consider in this paper ($\tau$ sufficiently remote from any quantum resonance), the localisation length is also nearly independent from the pulse-train period $\tau$, and therefore is solely a function of the effective interaction strength $P$ (or, generally, the interaction potential $V(\theta)$).
In Figure~\ref{fig.exponential} we show the population distribution of the angular momentum states for different parameter values.
One can clearly see the exponential  shape of the distribution (note the logarithmic scale).
In panel~(a) to~(c) only noise-free pulse trains are considered, and the interaction strength $P$ (a), the number of pulses $N$ (b), and the pulse train period $\tau$ (c) are varied.
One can clearly see that an increase of the interaction strength leads to an increase of the localisation length, but the number of pulses or the pulse train period have no influence (as long as the pulse train period is sufficiently detuned from a quantum resonance).

\begin{figure}
\includegraphics{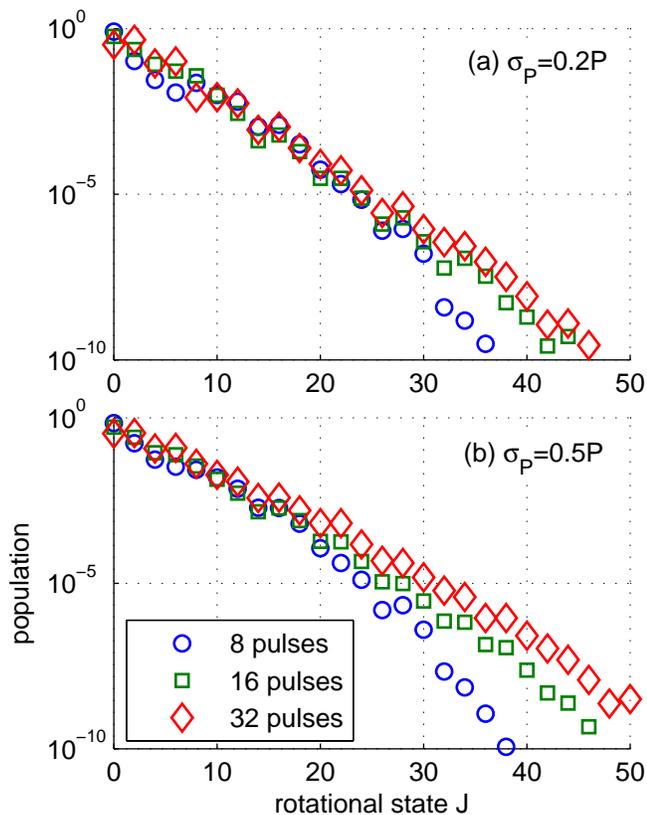}
\caption{\label{fig.exponentialnoise}(Colour online) Population of the even rotational states after a train of noisy $\delta$-pulses.
The pulse train period is $\tau=1$, the average effective interaction strength is $P=3$.
}
\end{figure}

Figure~\ref{fig.exponential}~(d) shows the influence of noise.
It can be seen that even for strong noise (diamonds), the population distribution still resembles an exponential curve.
Therefore, an exponential-like shape of the angular momentum population distribution is not an  unambiguous criterion for Anderson localisation.
However, the introduction of noise leads to a dependence of the width of the population distribution on the number of pulses, as is shown in Fig.~\ref{fig.exponentialnoise}.
For $\sigma_P=0.2P$ ($\sigma_P$ being the standard deviation of the interaction strength), the dependence is still very weak, but for stronger noise one can see the increase of the width of the distribution with $N$, which clearly demonstrates the destruction of Anderson localisation by strong noise.

The periodically kicked classical rotor can show chaotic dynamics with a diffusive growth of the energy; in the quantum mechanical regime this diffusion is suppressed due to Anderson localisation~\cite{casati79,fishman82,grempel84}.
If the product of the kick strength $P$ and the kicking period $\tau$ exceeds a critical value, the classical rotor exhibits unbounded chaotic motion.
The angular momentum undergoes (approximately) a random walk, and the mean square deviation of the angular momentum grows with the number of pulses.
This leads to a diffusive growth of the rotational energy.
For a quantum mechanical rotor, this diffusive growth is stopped after few pulses due to the Anderson localisation.
One can therefore see a linear growth of the energy till some critical number $N_{break}$ of pulses.
For later pulses, the energy oscillates, but does not grow on average~\cite{casati79}.
In general, $N_{break}$ increases with the effective interaction strength $P$~\cite{ott84}.

\begin{figure}
\includegraphics{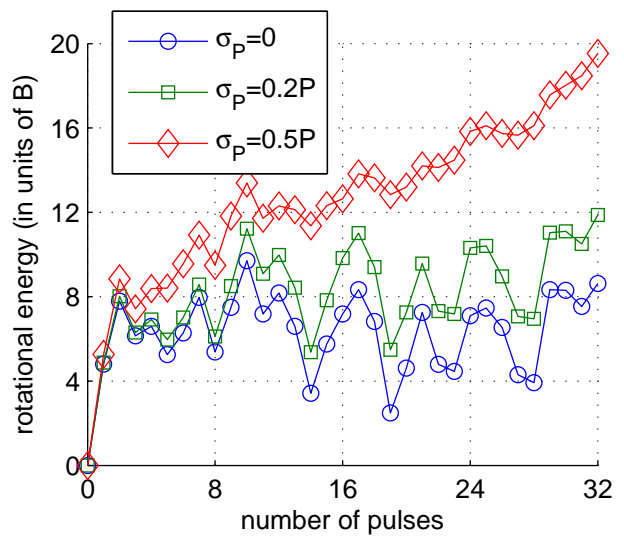}
\caption{\label{fig.suppression}(Colour online) Absorbed rotational energy for a periodically kicked quantum rotor.
Compared are the cases of a noiseless (circles) and a noisy pulse train (squares and diamonds).
The effective interaction strength is $P=3$, the pulse train period is $\tau=1$, and $\sigma_P$ is the standard deviation of the interaction strength noise.
}
\end{figure}

In Figure~\ref{fig.suppression} we show the suppression of the energy diffusion for a periodically kicked linear molecule.
Shown are the results for different noise strength: no noise (circles), moderate noise (squares, $\sigma_P=0.2P$), and strong noise (diamonds, $\sigma_P=0.5P$).
The interaction strength is chosen as $P=3$, which is comparable to laser intensities used in current experiments on laser alignment~\cite{cryan09}.
Regardless of the amount of noise, we can see a fast energy growth over the first two pulses.
After this initial phase, there is a diffusive growth of the energy for noisy trains, where the diffusion coefficient increases with the strength of the noise.
This is exactly what one can expect for a dynamically localised  periodically kicked rotor:
without noise, the energy diffusion is suppressed, but with increasing noise, the classical diffusion is eventually recovered~\cite{guarneri84,ott84}.
On the other hand, if the noise is too strong, the dynamics become dominated by the noise, but not by the underlying classical system.

Instead of measuring the population of all rotational levels after $N$ pulses, one can also observe localisation via the population of a single level measured as a function of $N$.
For a system localised at a level $J_0$, the population of this level must remain finite for $N\to\infty$, thus demonstrating non-zero survival probability.

\begin{figure}
\includegraphics{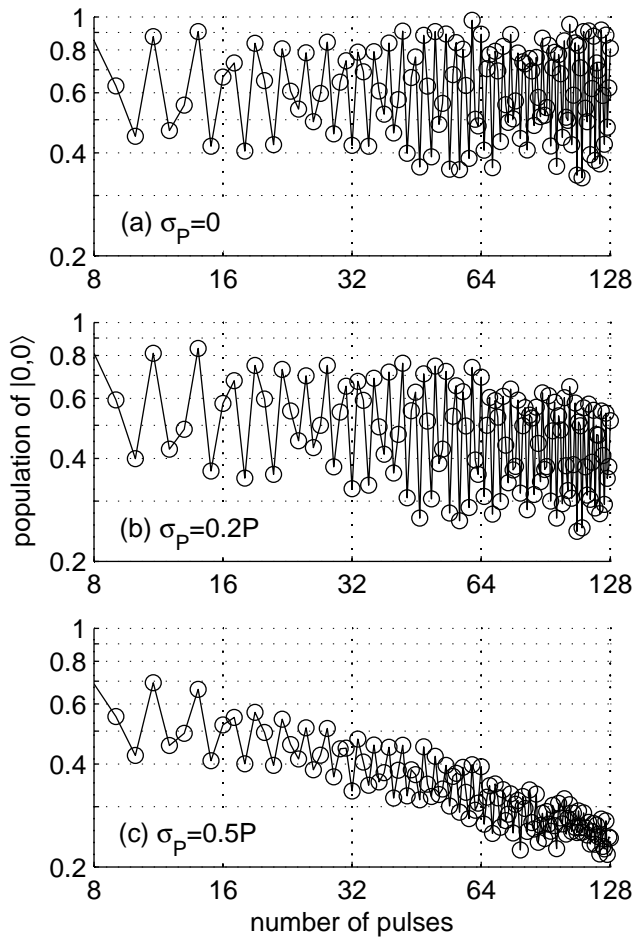}
\caption{\label{fig.finitepopulation} Population of the state $|0,0\rangle$ as a function of the number of pulses for trains with different standard deviations $\sigma_P$ of the interaction strength.
The pulse train period is $\tau=1$, and the effective interaction strength $P=3$.
Note the logarithmic scales.}
\end{figure}

In Figure~\ref{fig.finitepopulation} we show the population of the initial state (here: the ground state) for an off-resonant pulse train with $P=3$ as a function of the number of pulses.
For the noiseless pulse train the survival probability remains finite and oscillates around 0.6.
For a slightly noisy pulse train with $\sigma_P=0.2P$, one can see a slow decline of the survival probability for more than 32 pulses.
For a  pulse train with strong noisy, a clear power law decay (note the double logarithmic scale) is seen already after 32 pulses.

Note that a finite survival probability does not necessarily mean that the whole system is localised.
For a three-dimensional rotor like a linear molecule, the angular momentum $J$ can only take positive values, so one can expect ``surface effects'' for Eqs.~\eqref{eq.anderson}.
We observed that these ``surface effects'' lead to a trapping of some population in the low-lying initial $J$-state for fractional quantum resonances, although this system can still be excited unboundedly.
This issue will be discussed in more detail in a future publication.



\section{\label{sec.limitations}Considerations for an experiment}

In the proceeding section, we considered a rigid molecule at zero temperature interacting with ideal $\delta$-kicks.
We will now show that current experimental techniques are close enough to this idealised case to allow observation of Anderson localisation.

\subsection{Experimental constraints}

A long pulse duration can be a rather trivial cause for localisation in angular momentum space.
If the pulse is long compared to the relevant excitation periods, the interaction is adiabatic and no net excitation is observed after the pulse.
This prevents the excitation of higher angular momentum levels, leading to localisation of the molecule in angular momentum space.
In the following, we will refer to this localisation mechanism as adiabatic localisation, in contrast to Anderson localisation.
In order to exclude the adiabatic localisation effect, the pulse duration should be much shorter than the rotational period of the molecule.
For a molecule in the level $J$, this period is given as
\begin{equation}
t_{exc}(J) \approx \frac{2\pi I}{L} = \frac{t_{rev}}{J}\,,
\label{eq.diabatic}
\end{equation}
where $I$ is the moment of inertia, $L=\hbar J$ is the  angular momentum, and $t_{rev}=2\pi I/\hbar$ is the so-called rotational revival time, which in dimensionless units is $t_{rev}=2\pi$.
As an example, for $^{14}$N$_2$ the rotational period is approximately $8.4/J~\text{ps}$.

From Fig.~\ref{fig.exponential} we can see that the exponential localisation can already be very well observed if it covers the levels up to $J=20$.
Using Eq.~\eqref{eq.diabatic}, the excitation period of $J=20$ is $t_{exc}\approx0.05 t_{rev}$, so we may expect a pulse duration shorter than $0.01 t_{rev}$ to be sufficient for the clear observation of Anderson localisation.

\begin{figure}
\includegraphics{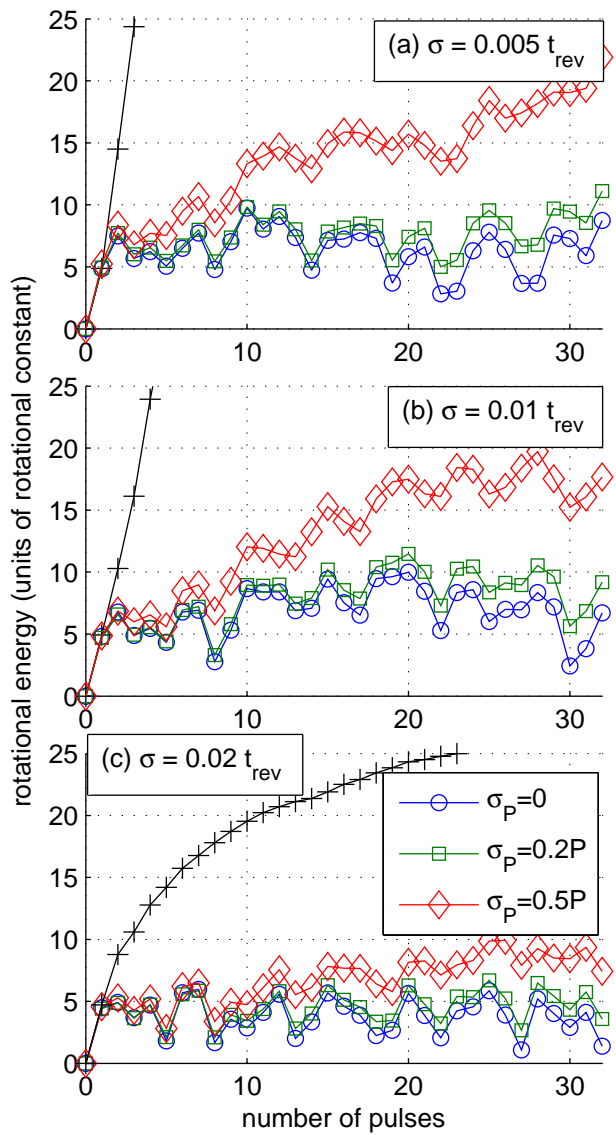}
\caption{\label{fig.suppression_finitepulse}(Colour online) Rotational energy as a function of the number of pulses for three different pulse durations.
Shown are the results for different standard deviations $\sigma_P$ of the kicking strength.
For comparison we also show the energy absorption of a corresponding classical rotor (black crosses).
The effective interaction strength is $P=3$ and the pulse train period is $\tau=1$.}
\end{figure}

In Figure~\ref{fig.suppression_finitepulse} we show the rotational energy of a kicked rotor for different pulse durations and different amounts of noise.
It can be seen that for pulse durations of $\sigma\leq0.01 t_{rev}$, indeed the suppression of diffusion can be observed as good as for a $\delta$-pulse (see Fig.~\ref{fig.suppression}).
For a slightly longer pulse ($\sigma=0.02 t_{rev}$, shown in Fig.~\ref{fig.suppression_finitepulse}~(c)) on the other hand, the adiabatic localisation is not negligible any more and leads to additional localisation.
A good test for adiabatic localisation is the comparison of a kicked quantum rotor to a kicked classical rotor.
The Anderson mechanism does not apply to a classical rotor, so the localisation seen for the classical rotor is only due to the adiabatic localisation.
In Figure~\ref{fig.suppression_finitepulse} we therefore also show the rotational energy of a kicked classical rotor (black crosses).
As can be seen, in all cases the quantum rotor is much stronger localised than the classical rotor, even for the rather long pulse duration of $\sigma=0.02 t_{rev}$.
This shows that even for long pulses, one may still observe dynamical localisation, although it is mixed with adiabatic localisation.

Up to now, we treated the molecules as rigid rotors, neglecting vibrational motions.
For diatomic molecules this is a well justified approximation:
Although the vibrations change the moment of inertia, the vibrational motion is much faster than the rotation, so one can average over the vibrations and arrive at an average moment of inertia.
The rotational levels including the vibrational effects are given as
\begin{equation}
E_{J,v}=B_eJ(J+1)-D_eJ^2(J+1)^2-\alpha_e(v+\frac{1}{2})J(J+1) \,.
\label{eq.levels_diatomic}
\end{equation}
Here, $B_e$ is the rotational constant of the molecule in its equilibrium configuration, and $v$ is the vibrational quantum number.
The second term in~\eqref{eq.levels_diatomic} accounts for centrifugal forces, i.e. bond stretching due to fast rotations, and the third term accounts for the change of the average bond distance in higher rotational levels.
Since the laser pulses under consideration do not couple different vibrational states, $v$ can be treated as a parameter and the rotational levels can be written as
\begin{equation}
E_{J}^{(v)}=B_vJ(J+1)-D_eJ^2(J+1)^2 \,,
\end{equation}
where $B_v=B_e+\alpha_e(v+1/2)$.
The condition for Anderson localisation is that the series~\eqref{eq.TJ},
\begin{equation}
T_J^{(\alpha)}=\tan\left[\frac{\tau}{2}(E_J-\omega_{\alpha})\right] \,,
\end{equation}
is random (see above).
Obviously, if this series is pseudo-random for a rigid rotor ($D_e=0$), it is such also for a non-rigid rotor with $D_e\neq0$.
Therefore, Anderson localisation can be observed in a non-rigid diatomic molecule as well as in a rigid one.

More complicated is the case of polyatomic linear molecules like carbon disulfide.
Here, only in the vibrational ground state the molecule can be regarded as linear.
Vibrational excitations, e.g. by thermal effects, can render the molecule to a symmetric-top rotor, strongly influencing the rotational dynamics~\cite{townes55}.
Although Anderson localisation in a symmetric top may be an interesting case as well, we  restrict ourselves to linear molecules here, and therefore do not consider polyatomic molecules in more detail.

\begin{figure}
\includegraphics{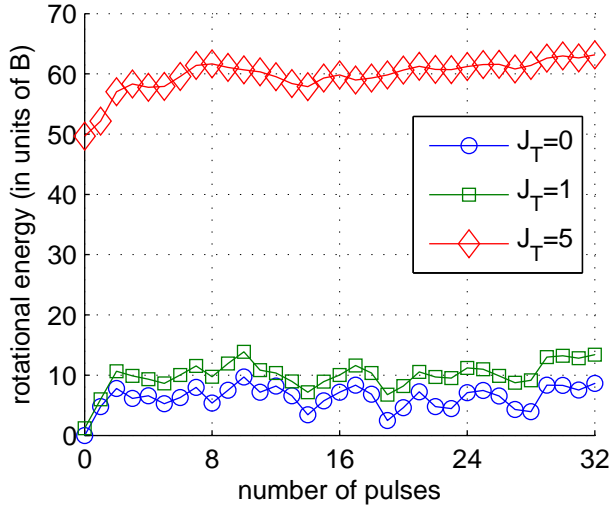}
\caption{\label{fig.temperature}(Colour online) Rotational energy of a rigid diatomic molecule as a function of the number of pulses for three different effective temperatures $J_T=\sqrt{T k_B/(2B)}$.
The effective interaction strength is $P=3$ and the pulse train period is $\tau=1$.}
\end{figure}

In order to investigate the influence of the temperature, it is convenient to introduce an effective thermal value of the angular momentum, $J_T=\sqrt{T k_B/(2B)}$, where $k_B$ is Boltzmann's constant.
For numerical purposes, we include the temperature by doing ensemble averaging over the initial state, weighting each result by its initial state's Boltzmann factor.
In Fig.~\ref{fig.temperature} we show the absorbed rotational energy for a rigid diatomic molecule at different temperatures, corresponding to $J_T=0,1,5$.
For nitrogen e.g. these temperatures correspond to $T\approx0~\textrm{K}, 6~\textrm{K}, 143~\textrm{K}$, respectively.
It can be seen that an increase of the temperature leads to an increase of the baseline and a slight smoothing of the oscillations, but not to a qualitative change of the energy absorption.

An additional effect of an increased temperature is vibrational excitation.
Since the laser does not couple different vibrational states, the thermal population of excited vibrational levels would not change the system in general.
However, it would lead to independent ensembles with a different rotational constant, which may disturb the observations.
In molecular beams, a rotational temperature of less than 10~K is generally reached.
However, the vibrational motion is only slightly cooled in a molecular beam.
This can be a complicating problem for heavy diatomic or polyatomic linear molecules.
For the latter, excitation of the bending modes may change them from linear rotors to symmetric tops~\cite{townes55}.
In this paper we therefore consider only light diatomic molecules, for which vibrational excitation is negligible.
E.g. for N$_2$ at room temperature, the population of the first excited vibrational state is only $10^{-5}$ of the ground state population.

\subsection{Example calculation}

To demonstrate that the proposed experiments are possible with current laser technology, we now show simulations for $^{14}$N$_2$ molecules, interacting with laser pulse trains with experimentally feasible parameters.
The pulse duration is chosen as $\sigma=0.005 t_{rev}\approx40~\mathrm{fs}$, and the peak intensity is $I_0=3\cdot10^{13}~\mathrm{W}/\mathrm{cm}^2$, which corresponds to $P=2.9$.
These pulse parameters are close to the ones used by Cryan \textit{et al.}~\cite{cryan09} for laser alignment of nitrogen.
As initial rotational temperature we chose $T=8~\mathrm{K}$.
For the vibrations, we assume that all population is in the ground state, which is well justified even if the vibrational temperature was equal to room temperature (the vibrational frequency of $^{14}$N$_2$ is $2359~\mathrm{cm}^{-1}$).
We compare a slightly noisy pulse train with $\sigma_P=0.2P$ and a pulse train with stronger noise of $\sigma_P=0.5P$.

\begin{figure}
\includegraphics{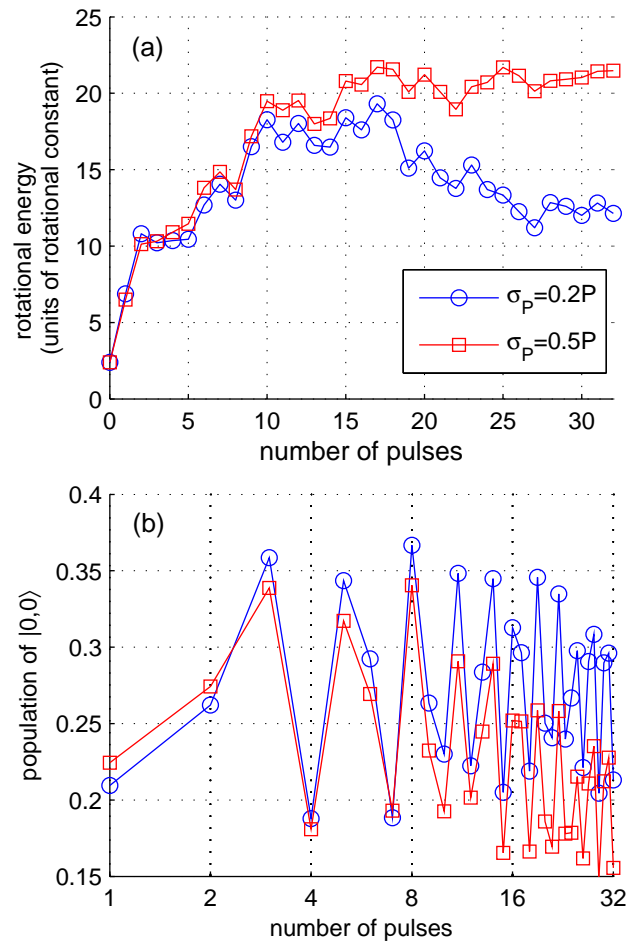}
\caption{\label{fig.nitrogen}(Colour online) Absorbed rotational energy (a) and population of ground rotational state (b) for $^{14}$N$_2$ molecules interacting with a train of linearly polarised laser pulses, shown for weak (blue circles) and strong (red squares) noise.
The pulse duration is $\sigma=0.005 t_{rev}\approx40~\text{fs}$, and the peak intensity is $I_0=3\cdot10^{13}~\mathrm{W}/\mathrm{cm}^2$, corresponding to an effective interaction strength of $P=2.9$ for every pulse.
The initial rotational temperature of the molecules is $T=8~\mathrm{K}$.
In panel (a), the rotational energy for the slightly noisy case seems to decrease after 16 pulses, but it actually starts to oscillate.
These oscillations could be seen if a train of more than 32 pulses was used.
Note the logarithmic scale in panel (b).}
\end{figure}

\begin{figure}
\includegraphics{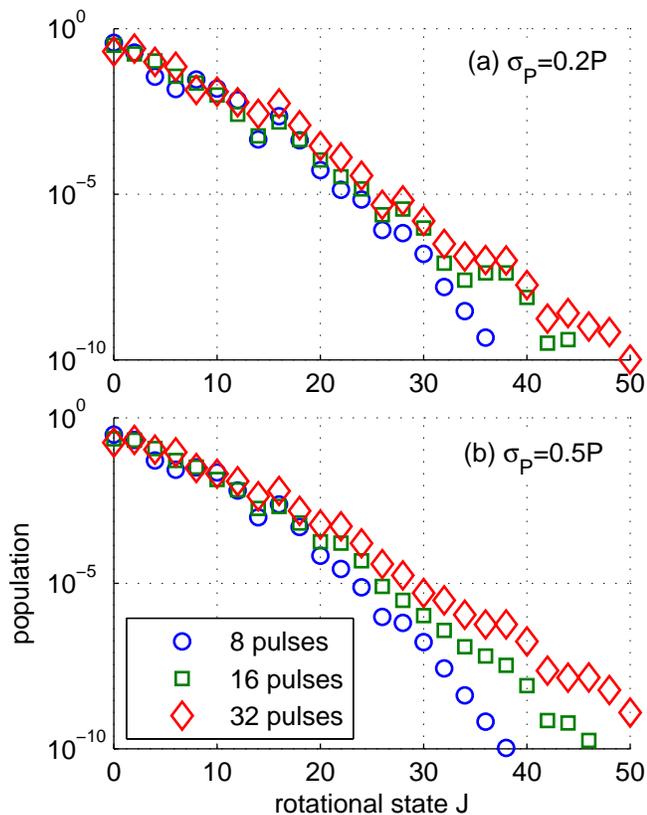}
\caption{\label{fig.exponential_nitrogen}(Colour online) Population of the even rotational levels of $^{14}$N$_2$ after interaction with a train of up to 32 laser pulses.
The pulse duration is $\sigma=0.005 t_{rev}\approx40~\text{fs}$, and the peak intensity is $I_0=3\cdot10^{13}~\mathrm{W}/\mathrm{cm}^2$, corresponding to an effective interaction strength of $P=2.9$ for every pulse.
The initial rotational temperature of the molecules is $T=8~\mathrm{K}$.
}
\end{figure}

In Fig.~\ref{fig.nitrogen}~(a), we show the rotational energy as a function of the number of pulses.
One can clearly see a suppression of the diffusion.
Moreover, this suppression is weakened when the noise is increased.
Note that the rotational energy for the slightly noisy case seems to decrease after 16 pulses, but it actually starts to oscillate.
These oscillations could be seen more clearly if a train of more than 32 pulses was used.
The population of the rotational ground state [Fig.~\ref{fig.nitrogen}~(b)] shows a very slow decay for the weakly noisy train, and a faster decay for the strongly noisy train.
In Fig.~\ref{fig.exponential_nitrogen} we finally plot the population of the rotational levels.
Since the even and odd rotational levels are independent ensembles (the interaction couples only $\Delta J=\pm2$), we show only the even levels for clarity.
One can clearly see the exponential decay of the population with increasing $J$.
For the pulse train with weak noise, the population distribution is almost independent of the number of pulses.
With increased noise, the population distribution and in particular its width in $J$-space becomes more dependent on the number of pulses.

Concluding, one can see clear signs of localisation in nitrogen molecules interacting with state-of-the-art laser pulse trains.
In particular, one can see a weakening of the localisation when introducing noise, which is an indication that the Anderson mechanism is causing the localisation.



\section{\label{sec.discussion}Discussion and conclusion}

The arguments presented in this work show that current laser technology used for molecular alignment is sufficient for inducing Anderson-like localisation of kicked molecules in the angular momentum space.
In the experiment by Cryan~\textit{et al.}~\cite{cryan09}, a train of eight pulses with duration of 50~fs and peak intensity of $36~\text{TW}/\text{cm}^2$ was used to induce strong molecular alignment.
In this work we have shown that in the very same experiment one could have induced Anderson-like localisation, if the pulse train period would not have been chosen as exactly the rotational revival time, but would have been generally detuned from it.

The main remaining challenge is not inducing the Anderson localisation, but  observing  it.
We presented and analysed different indicative signatures of the dynamical localization in the laser-kicked molecules.
The hardest to observe, but the most direct one is the exponentially falling-down distribution of the population of the molecular rotational states.
Our simulations indicate that it will be necessary to measure the population of several tens of rotational states with the population values over the range of at least two or three orders of magnitude.
However, instead of measuring the population of all rotational states, one may opt for monitoring the population just of the ground state and demonstrate that it has a finite survival probability.
For this purpose, much less accurate measurement of the rotational state population is sufficient.
Moreover, suppression of the energy diffusion due to the Anderson localisation causes molecules to stop absorbing the energy after the first few pulses.
This might be observed as an increased transparency of the medium.

Measurements on molecular alignment induced by laser pulse trains may provide the needed information for detecting the dynamical Anderson localisation.
The alignment is usually quantified by the expectation value $\langle \cos^2\theta \rangle (t)$, called alignment factor, and there are multiple techniques to measure it~\cite{ourreview,ohshimareview,stapelfeldt03}.
Time analysis of the time-dependent alignment signals as measured in the experiment by Cryan~\textit{et al.}~\cite{cryan09} could provide data on the population of rotational levels with the help of reconstruction procedures similar to the one described in~\cite{ezra}.

A different approach, may rely on the measurement of the time-average value of the alignment factor, also called population alignment.
For a molecule in thermal equilibrium, the population alignment is exactly $1/3$.
On the other hand, under diffusive conditions, the angular momentum $J$ is "unlimitedly" increasing by the periodic train of linearly polarized pulses, whilst the projection quantum number $M$ remains constant.
Under these conditions, the molecules are finally rotating in vertical planes (containing the polarization vector), and the time-averaged alignment factor asymptotically approaches the value of $1/2$.
If the measured population alignment saturates at the level below $0.5$, this may be an indicator of the suppression of chaotic diffusion due to the Anderson localisation.

The results presented here show that the goal of observing Anderson-like localisation in a molecular system can be achieved by using proven laser technology.
We hope that this work will encourage corresponding experiments in the near future.



\begin{acknowledgements}

We appreciate useful discussions with Paul Brumer, Phil Bucksbaum, Valery Milner, Yehiam Prior, Evgeny Shapiro, Moshe Shapiro, Uzy Smilansky, Michael Spanner, and Sergey Zhdanovich.
This work was supported in part by the National Science Foundation (Grant No. PHYS-1066293), the US-Israel Binational Science Foundation (Grant No. 2010132), the Israel Science Foundation (Grant No. 601/10), the Deutsche Forschungsgemeinschaft (Grant No. LE 2138/2-1), and the Minerva Center of Nonlinear Physics of Complex Systems.
IA acknowledges support as the Patricia Elman Bildner Professorial Chair.
SF acknowledges support as the Shlomo Kaplansky Academic Chair and appreciates the hospitality of the Aspen Center for Physics.
This research is made possible in part by the historic generosity of the Harold Perlman Family.
\end{acknowledgements}


%

\end{document}